\documentclass[letterpaper,12pt]{article}   
\usepackage{osajnl2} 
\usepackage[draft]{hyperref} 
\usepackage{amsmath,amssymb,graphicx,subfigure}

\begin{document}

\title{Angular instability due to radiation pressure in the LIGO gravitational wave detector}


\author{E. Hirose,$ ^{1,4,*} $ K. Kawabe,$ ^{2} $ D. Sigg,$ ^{2} $ R. Adhikari,$ ^{3} $ and P.R. Saulson$ ^{1} $}
\address{$^1$Department of Physics, Syracuse University, \\ Syracuse, NY 13244, USA}
\address{$^2$LIGO Hanford Observatory, \\ 127124 North Route 10, Richland, WA 99354, USA}
\address{$^3$LIGO Laboratory, California Institute of Technology, \\MS 18-34, Pasadena, CA 91125, USA}
\address{$^4$Currently with Harvard-Smithsonian Center for Astrophysics, \\60 Garden St. MS 63, Cambridge, MA 02138 USA}
\address{$^*$Corresponding author: ehirose@cfa.harvard.edu}

\begin{abstract}
We observed the effect of radiation pressure on the angular sensing and control system of the Laser Interferometer Gravitational-Wave Observatory (LIGO) interferometer's core optics at LIGO Hanford Observatory.  This is the first measurement of this effect in a complete gravitational wave interferometer.  Only one of the two angular modes survives with feedback control, since the other mode is suppressed when the control gain is sufficiently large.  We developed a mathematical model to understand the physics of the system. This model matches well the dynamics that we observe.\\
\end{abstract}

\ocis{000.2780, 120.3180, 120.2230, 220.1140.}

\maketitle 



\section{Introduction}

Gravitational wave detection technology has made great strides in recent years, and more progress is expected in the near future. In the U.S., the Laser Interferometer Gravitational-Wave Observatory (LIGO) project has constructed several large interferometers, two at Hanford WA and another one at Livingston LA. \cite{BarryRai} These initial LIGO interferometers were used to make observations of unprecedented sensitivity during the period November 2005 through September 2007, LIGO's fifth science run, during which one year's worth of triple coincident data were collected at an rms strain sensitivity of about $ 10^{-21} $ \cite{Waldman,Sigg1,Kawabe}. The LIGO Scientific Collaboration (LSC) is now completing the analysis of these data.  Operation of LIGO and analysis of gravitational wave data are closely coordinated with two European projects, GEO (a British-German collaboration with an $ 0.6 $ km interferometer near Hannover \cite{GEO}) and Virgo (a French-Italian collaboration with a $ 3 $ km interferometer near Pisa \cite{VIRGO}).\\

\noindent Striking as this technological progress is, it is widely expected that sensitivity improvement of about a factor of ten in strain amplitude will be required before observations of gravitational waves become routine. Fortunately, the technology to make the required improvement is in hand. For LIGO, those improvements will be embodied in Advanced LIGO, now under construction and expected to begin operations around 2015 \cite{AdvLIGO}. In the meantime, one additional data run will be carried out with an improved version of the initial LIGO hardware. This incremental improvement, called Enhanced LIGO, will run during 2009 and 2010.

\section{Gravitational wave interferometers}

Gravitational waves interact with matter by causing a distinctive strain (fractional separation change) pattern in sets of test masses that are free to move. The distinctive pattern is that the sign of the strain is opposite in two perpendicular directions in the plane normal to the wave's propagation direction. For example, a wave propagating in the $ z $ direction causes equal and opposite strains in directions $ x $ and $ y $. This means that a well designed Michelson interferometer can be a good sensor of gravitational waves. A properly aligned wave traveling in the vertical direction would cause simultaneous-but-opposite changes in the lengths of the two arms, thus yielding a phase difference between the light from the two arms that can be observed when the light recombines at the beamsplitter. \\

\noindent Such a Michelson interferometer needs to include the following features: mirrors that are free to move in response to the wave (playing the role of the test masses), isolation of those mirrors from other causes of motion other than gravitational waves (such as the vibration of the laboratory floor), large length (so that strain is converted to a sufficiently large relative motion between the mirrors) and sufficient phase sensitivity to allow the motions to be observed. The fact that strain sensitivity of $ 10^{-21} $ has still not yielded a detection shows that these demands are severe.\\

\noindent LIGO has addressed these challenges as follows. \cite{BarryRai}  (Other interferometers have solutions that are similar in most features but differ in details.)  LIGO's interferometers have arms that are 4 kilometers in length. (The second of the two interferometers at LIGO Hanford Observatory is only 2 kilometers long.) The 10 kg mirrors are suspended as carefully engineered pendulums, which are in turn connected to multi-stage seismic isolation systems. A high power Nd:YAG laser (with a power of 10 W) is used to provide sufficient optical power to reduce the phase noise due to photon shot noise. The simple Michelson interferometer configuration is augmented by several features that substantially enhance efficiency. Each arm is made up of a Fabry-Perot cavity held on resonance, thus amplifying the phase shift by a factor of $ \mathcal{F}/ \pi $ (the cavity's finesse $ \mathcal{F} $ is about 200). The effective laser power is enhanced by an additional factor of about 50 by placing a partly transmitting mirror between the laser and the rest of the interferometer; the combination of this recycling mirror and the rest of the interferometer forms an additional Fabry-Perot cavity that is held on resonance. The whole system is enclosed in a vacuum system held at $ 10^{-8} $ torr. (See the schematic diagram in Figure \ref{fig:extraction}.)\\

\section{Role of radiation pressure}

Consideration of the LIGO design reveals that radiation pressure cannot be neglected in its operation. Light arrives at the recycling mirror with a power of 5 watts, but inboard of the recycling mirror the effective power level hitting the interferometer's beamsplitter is 250 watts. Inside the resonant cavities that make up each of the long Michelson arms, the power level is 12 kilowatts. Thus, the radiation pressure on each of the arm mirrors is $ 10^{-4} $ newtons. Incident on a 10 kg mirror held in a pendulum suspension with resonant frequency of 0.5 Hz, this force would cause a displacement of $ 10^{-6} $ m. Of course, a Michelson interferometer with suspended mirrors only functions properly with a control system to hold the mirrors at the proper separations and alignment; this control system acts to counteract the radiation pressure force.  A coupling to longitudinal motion of the mirrors has been experimentally demonstrated by Nergis Mavalvala et. al \cite{Nergis2}.\\

\noindent In 2006, Sidles and Sigg \cite{SiggSidles} pointed out that torques due to radiation pressure can cause an interesting problem in interferometers with suspended mirrors. They showed that the angular coupling between the mirrors due to radiation pressure is best understood by considering the dynamics of a cavity as a set of normal modes. At large enough radiation pressures, one of these modes can become unstable.  Although LIGO detectors control not only longitudinal motion but also angular motion of the mirros, the servo bandwidth cannot be arbitrarily turned up due to noise coupling into the gravitational wave output. \\

\noindent In this paper, we first review the theory developed by Sidles and Sigg, giving a heuristic understanding of the instability that they predicted. We point out that initial LIGO operated successfully at power levels that would be expected to cause angular instability, if one made a naive application of their theory. In the next section, we present a measurement of the angular response to torque applied to a LIGO mirror, made under normal operating conditions (and at other light power levels as well.) The mirror response shows features different from those expected by the theory as presented in reference \cite{SiggSidles}. \\

\noindent The difference from the simple theory is not surprising, since a control system acts on the mirrors of the interferometer. In order to derive the expected response, we modeled the combined mechanical-optical-control system using Matlab and Simulink \cite{Mathworks}. Section 6 of the paper describes this model, and shows how our measurements agree with this more complete theory of the dynamics of the system. In particular, we can now understand the stability of initial LIGO at power levels at which instability was expected in an uncontrolled interferometer. Our measurements show that, inside the control system, the mirror's angular transfer function is indeed that of an unstable harmonic oscillator.\\

\noindent In the final section of the paper, we investigate how much the light power can be increased before the present control system would be overwhelmed by the radiation pressure instability.\\

\section{Review of Sigg-Sidles instability}

\begin{figure}[!h]
  \centering
\includegraphics[width=0.9\textwidth]{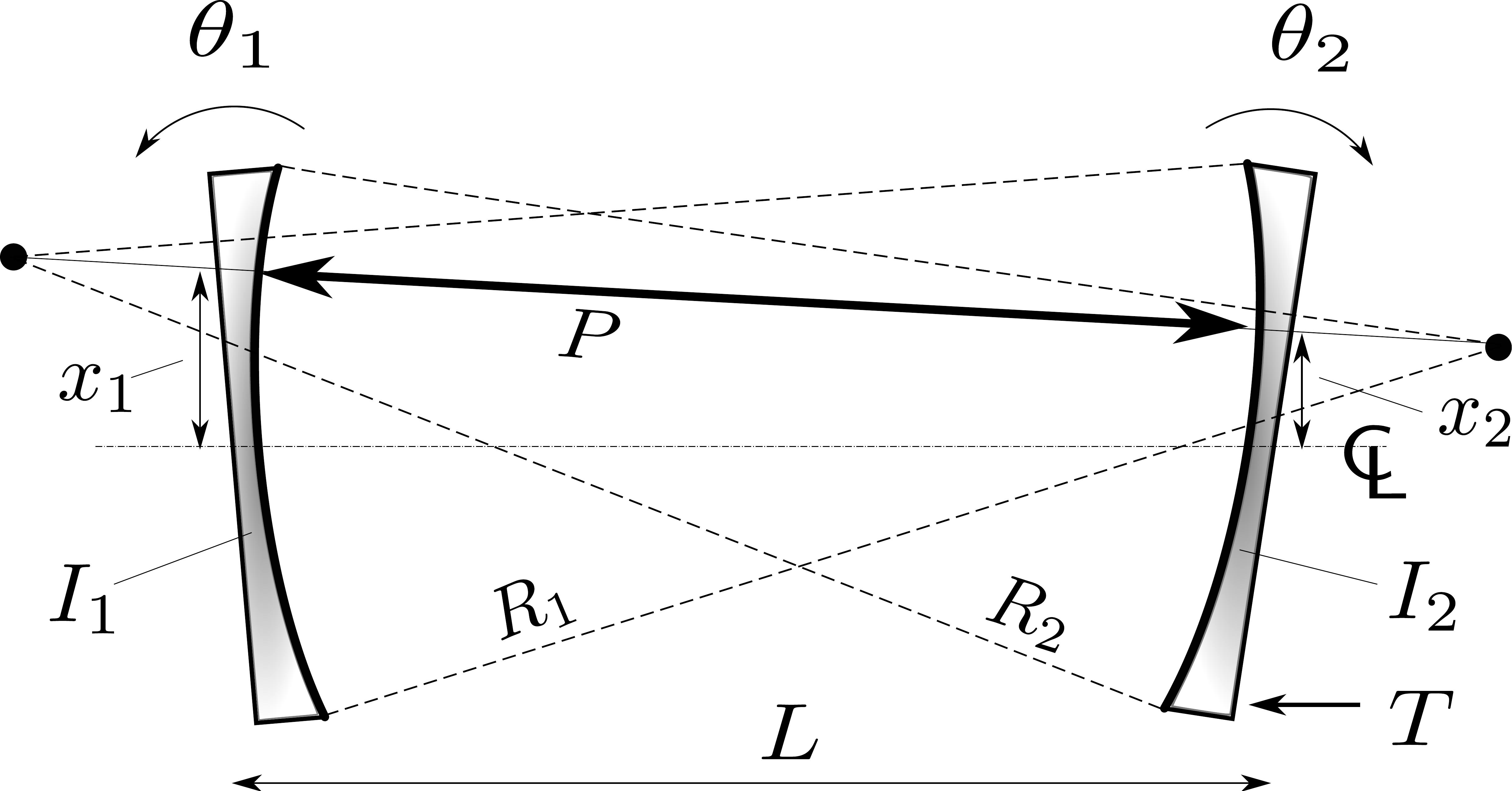} 
  \caption{Two mirrors are coupled by radiation pressure caused by the power of the laser beam $ P $, pushing at distance $ x_{1} $ and $ x_{2} $ away from the center line when the mirrors are tilted at angles of $ \theta_{1} $, $ \theta_{2} $ respectively.  $ I_{1} $, $ I_{2} $ are the moments of inertia of the mirrors.  $ R_{1} $, $ R_{2} $, $ L $ are the radii of curvature of the two mirrors and the distance between the mirrors.  $ T $ is an external torque applied to one of the mirrors to study the system.  The sketch can be viewed as either yaw motion or pitch motion.}
  \label{fig:tilted_resonator}
\end{figure}

Consider an optical resonator that consists of two suspended mirrors and laser light which resonates between the mirrors (Figure \ref{fig:tilted_resonator}).  The sketch shows two mirrors, which have very large radius of curvature (larger than the distance between two mirrors in LIGO detectors); the optical axis is the line connecting the centers of curvature of the mirrors.  The two mirrors feel a natural restoring force that comes from the wire loop suspending them from the suspension system.  \\

\noindent When the cavity is resonant, radiation pressure from the laser in the cavity can become important.  With no light, each mirror exhibits independent torsional oscillations.  When the cavity is filled with light, there are two coupled modes of the opto-mechanical system.  (This description can be applied either to the yaw mode or the pitch mode of the mirrors in the cavity. The measurements described below were made in the yaw mode.) \\

\noindent Suppose that the mirrors oscillate in such a way that the the sign of mirror angle $ \theta_{1}, \theta_{2} $ is either $ \left( +, - \right)  $ or $ \left( -, + \right)  $.  In this situation, radiation pressure works so as to enhance the original restoring force by pushing the mirrors back to the original position as shown in Figure 2, which means the mirrors experience a stronger restoring torque than that due to the mechanical restoring torque alone.  Thus, the eigenfrequency of this mode will be higher than the original uncoupled pendulum natural frequency.  \\

\begin{figure}[t]
  \centering
\includegraphics[width=0.7\textwidth]{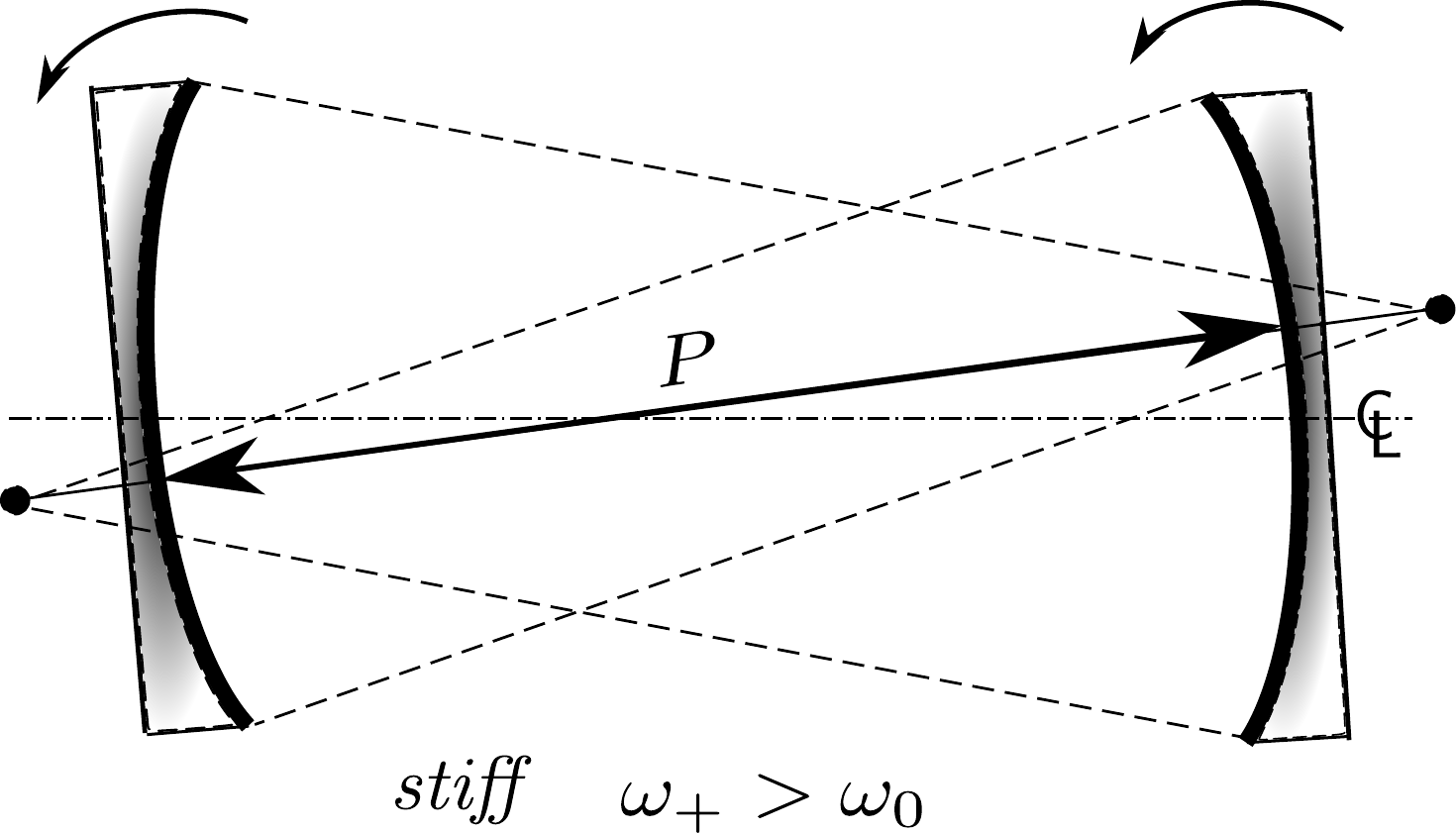} 
  \caption[The stiff mode of angular motion in an optical resonator]%
  {Stiff mode of angular motion in optical resonator.  Radiation pressure works to enhance the mechanical restoring force.}
  \label{fig:stiff}
\end{figure}

\noindent On the other hand, suppose that both mirrors tilt in such a way that the sign of mirror angle  $ \theta_{1}, \theta_{2} $ is either $ \left( +, + \right)  $ or $ \left( -, - \right)  $.  In that situation, the radiation pressure works against the original restoring force.  In this case, if the tilt angle becomes larger, the beam spot will move as shown in Figure 3, which means that the net restoring force will be smaller than that due to the mechanical restoring force alone.  When the power inside the optical resonator exceeds a critical value, the net restoring torque will become negative. \\

\begin{figure}[t]
  \centering
\includegraphics[width=0.7\textwidth]{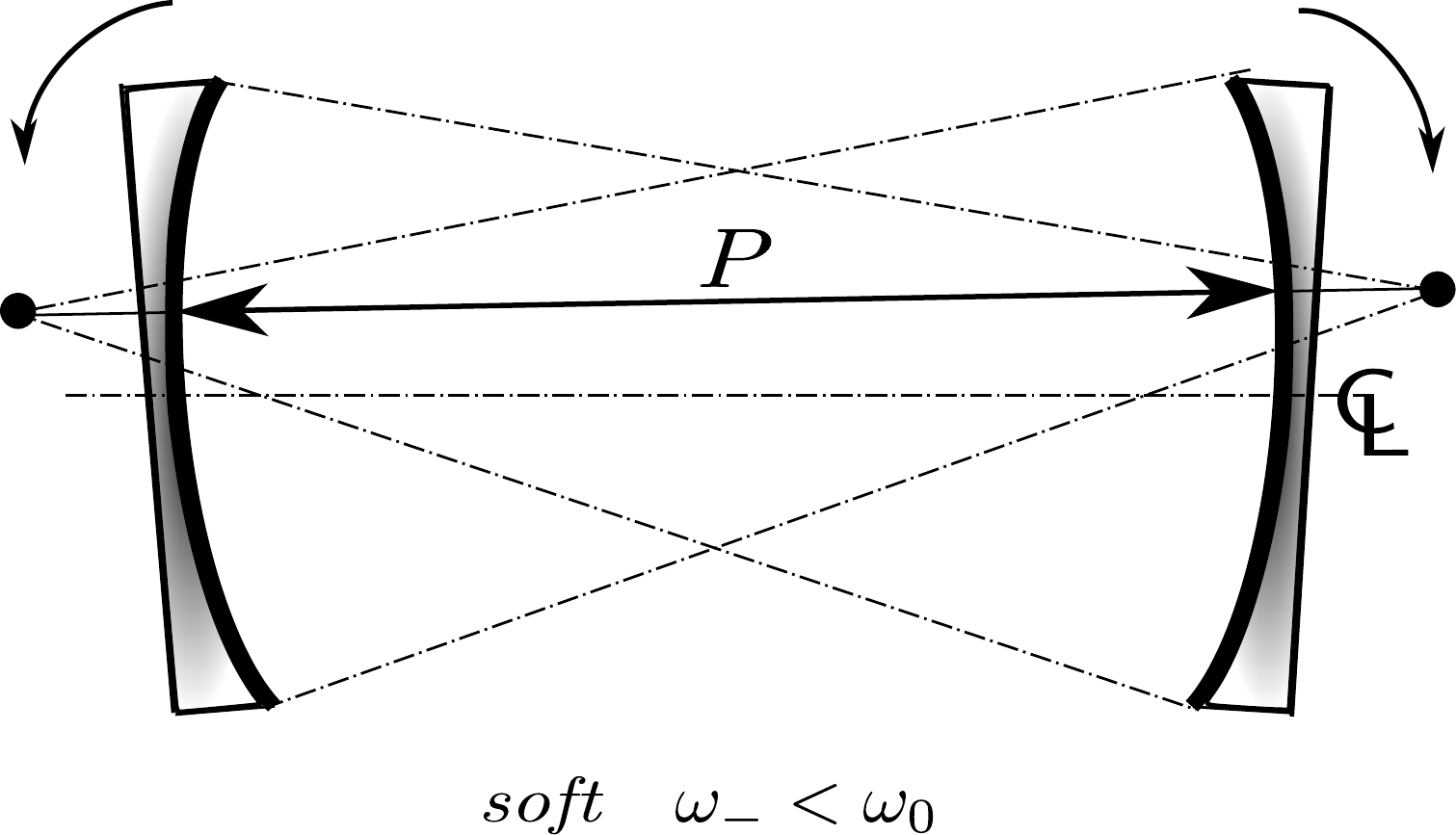} 
  \caption[The soft mode of angular motion in optical resonator]%
  {Soft mode of angular motion in optical resonator.  Radiation pressure works against the restoring force from the wire, and if the power exceeds a critical value, then this mode is unstable.}
    \label{fig:soft}
\end{figure}

\noindent In order to investigate the dynamics of this system, we can apply an external torque to one of the mirrors and determine the transfer function of the mirror's angle to the applied torque.  The normalized transfer function will be 

\begin{eqnarray}
H(s) = \frac{\theta_{2}}{T} \sim \frac{-\left(s^{2}+ \omega_{z}^{2} \right)}{\left(s^{2}+ \omega_{-}^{2}\right) \left(s^{2}+ \omega_{+}^{2}\right) }.
\label{eq:H_of_s}
\end{eqnarray}

\noindent The two pairs of poles $ \omega_{\pm} $ and the pair of zeros $ \omega_{z} $ are given by

\begin{eqnarray}
\omega_{\pm}^{2} &=& \omega_{0}^{2} + \frac{PL}{Ic}\left[ \frac{-\left( g_{1}+g_{2}\right) \pm \sqrt{4+\left( g_{1}-g_{2}\right) ^{2}}}{1-g_{1}g_{2}}\right], \\
\omega_{z}^{2} &=& \omega_{0}^2 - \frac{2PL}{cI}\frac{g_{2}}{1-g_{1}g_{2}},
\end{eqnarray}

\noindent where $ \omega_{0} $ is the mechanical resonant frequency of the torsion pendulum, $ P $ is the laser power inside the cavity, $ L $ is the length of the cavity, $ I $ is the moment of inertia of the mirror, and $ c $ is the speed of light. The factors $ g_{1} $ and $ g_{2} $ are the $ g $ parameters of the cavity, defined by $ g_{1,2} = 1-L \diagup R_{1,2} $ where $ R_{1} $ and $ R_{2} $ are the radii of curvature of the mirrors.  \\

\noindent Here is how one can have an intuitive justification of the transfer function, which describes the process of shaking one of the mirrors and observing the response of that mirror.  Since there are two angular modes of the coupled system, one will see resonances at the two different frequencies which correspond to the two pairs of poles $ \omega_{\pm} $.  Given the phase relations between these two modes, there will be a frequency at which the mirror does not move at all.  This frequency corresponds to a pair of zeros $ \omega_{z} $.  If we observe the other mirror, however, the pair of zeros does not show up.  \\

\begin{figure}[!h]
  \centering
\includegraphics[width=0.9\textwidth]{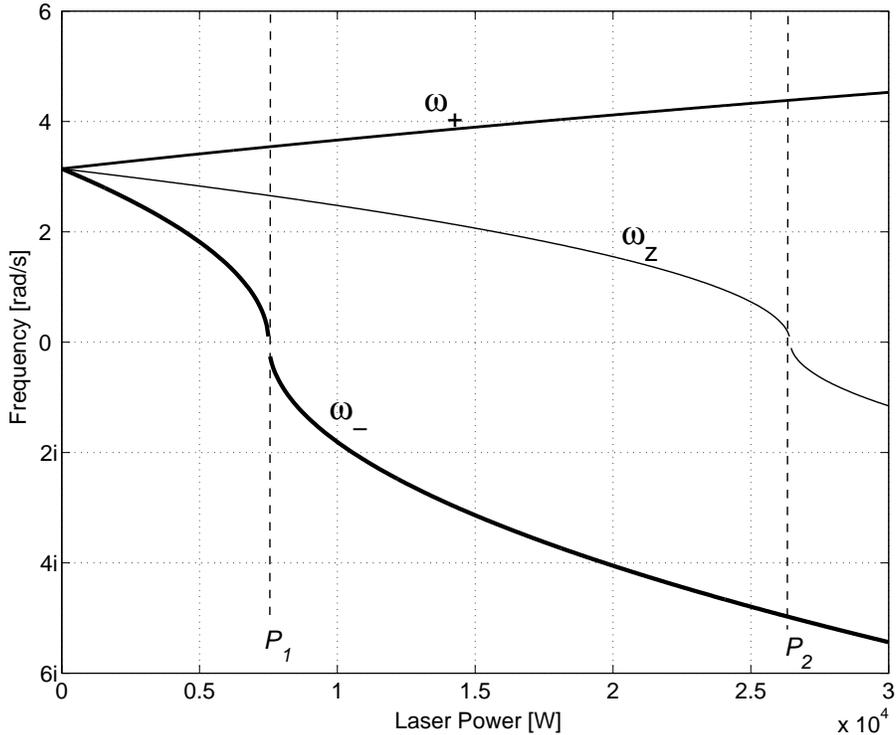} 
  \caption{Angular frequencies of poles and zeros as a function cavity laser power, according to Equations (4.1) - (4.3).  As $ P $ increases, $ \omega_{+} $ increases, while $ \omega_{-} $ and $ \omega_{z} $ decrease as $ P $ increases and become negative above $ P_{1} $, $ P_{2} $ respectively.}
\end{figure}

\noindent As $ P $ increases, $ \omega_{+} $ increases, while $ \omega_{-} $ and $ \omega_{z} $ decrease. (See Figure 4.)  Eventually, both $ \omega_{-}^2 $ and $ \omega_{z}^2 $ become negative, and the corresponding poles and zeros become real.  $ P_{1} $ and $ P_{2} $ in Figure 4 are the powers that give $ \omega_{-}^{2} = 0 $ and $ \omega_{z}^{2} = 0 $, respectively.  For initial LIGO, they are roughly $ P_{1} = 7.5 $ kW and $ P_{2} = 26.4 $ kW.  A set of Bode diagrams are shown in Figure 5, and a pole map of (\ref{eq:H_of_s}) is shown in Figure 6.\\

\begin{figure}[!h]
\centering
\subfigure [$ P = 0 $]{\includegraphics [width=0.48\textwidth]{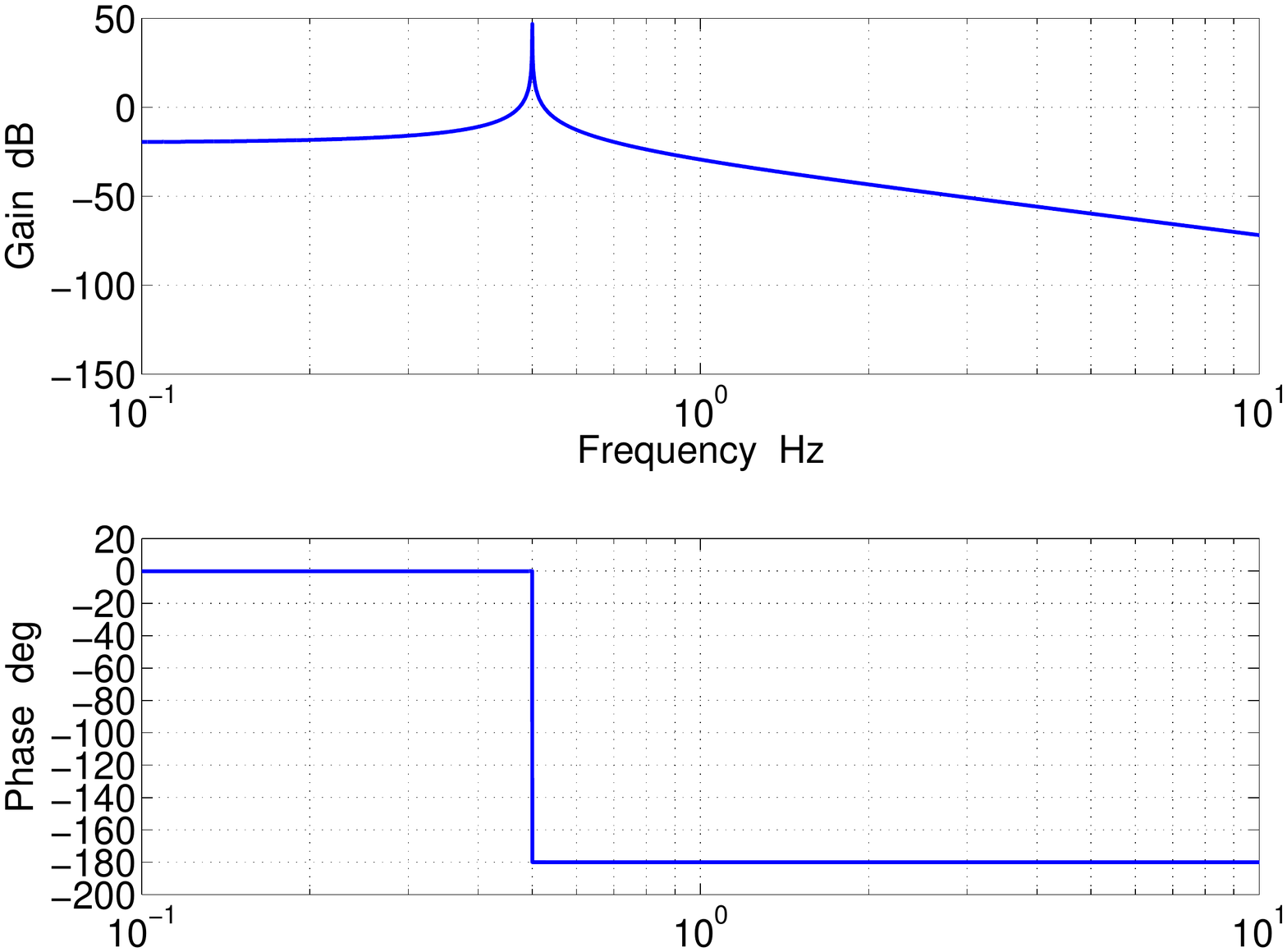}}
\subfigure [$ 0 < P < P_{1} $]{\includegraphics [width=0.48\textwidth]{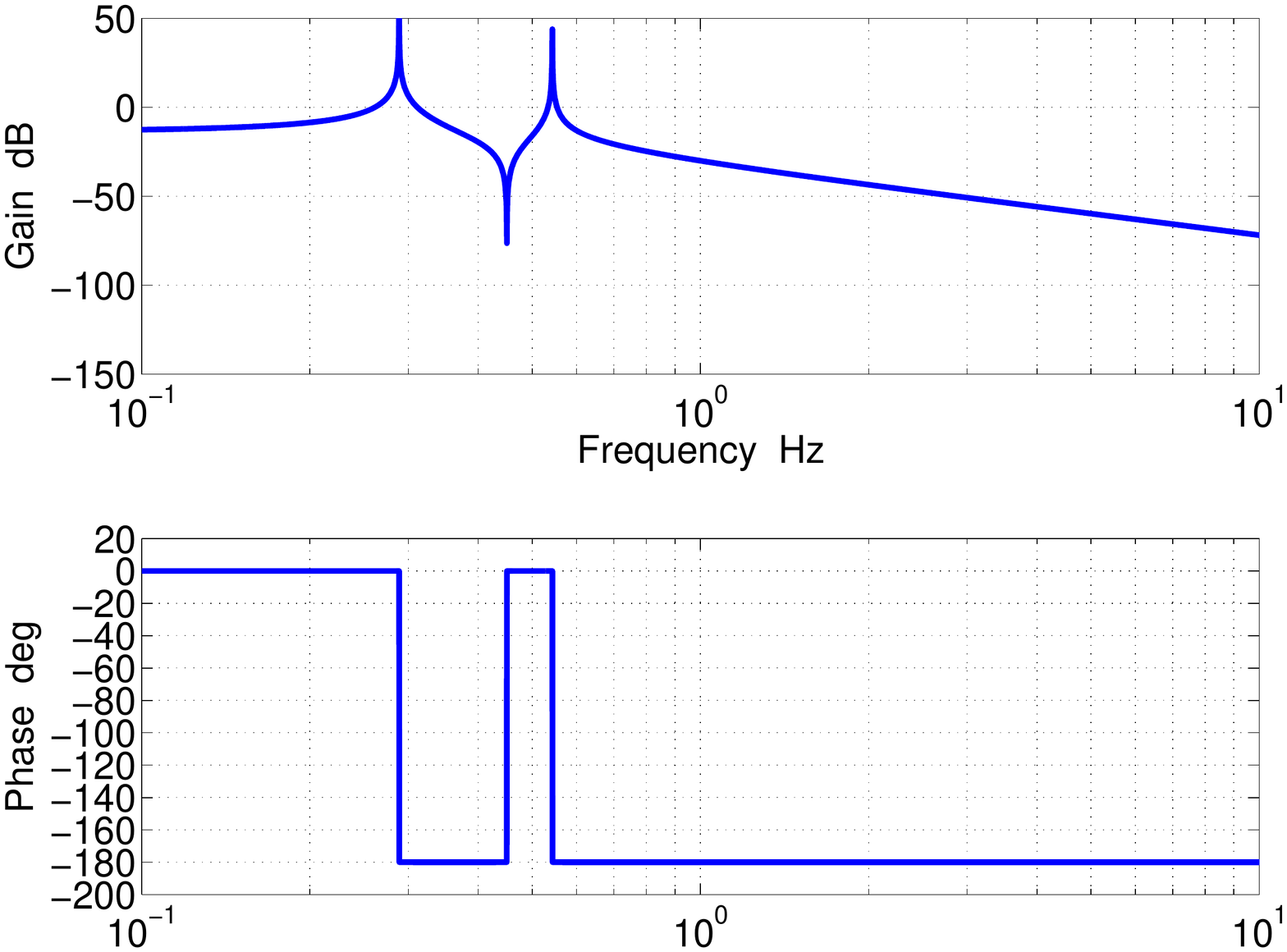}}
\subfigure [$ P_{1} < P < P_{2} $]{\includegraphics [width=0.48\textwidth]{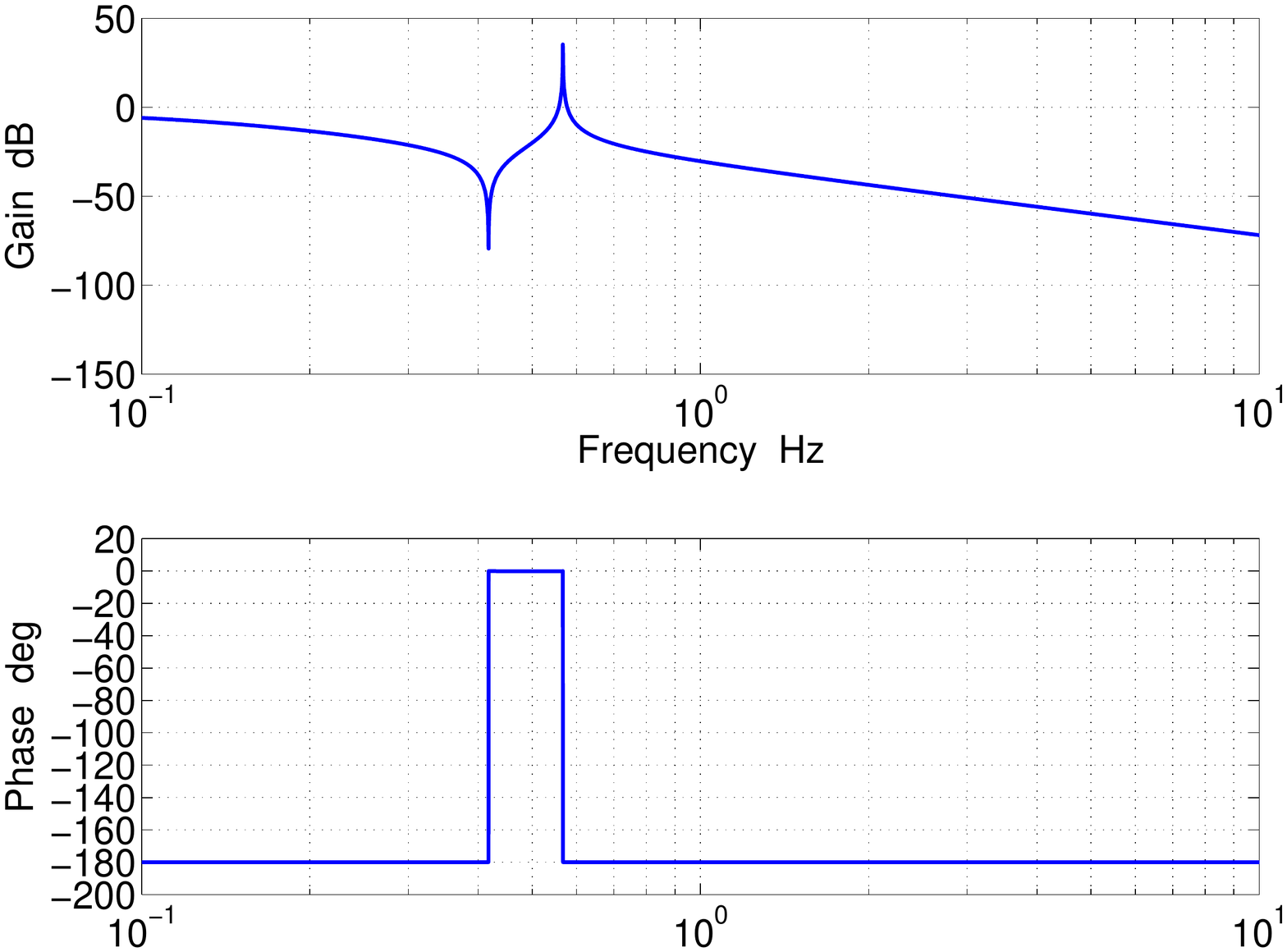}}
\subfigure [$ P_{2} < P $]{\includegraphics [width=0.48\textwidth]{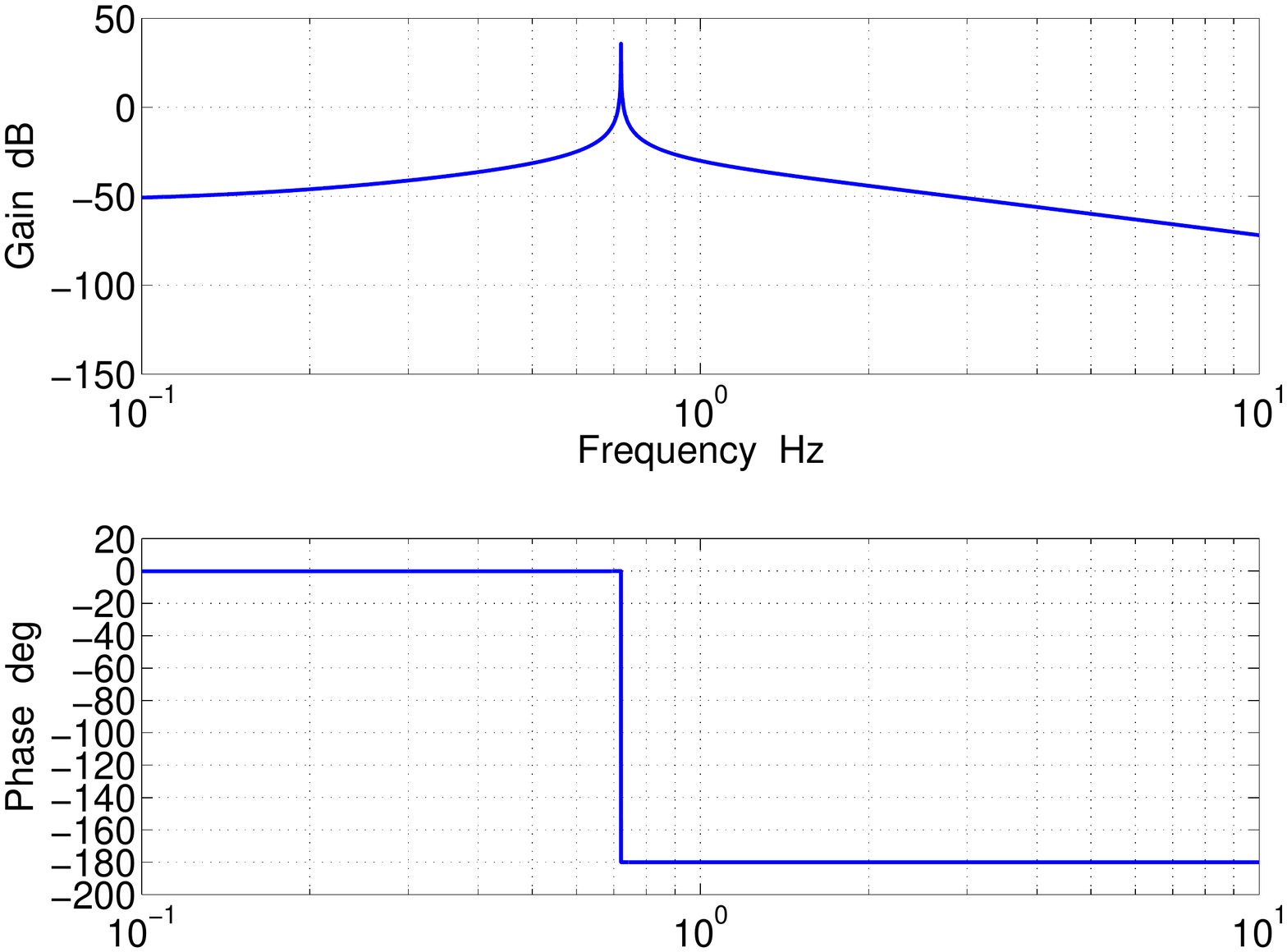}}
\caption [Bode plot]%
{Bode plot of the uncontrolled two-mirror system at various powers.  (a) $ P = 0 $, (b) $ 0 < P < P_{1} $, (c) $ P_{1} < P < P_{2} $, (d) $ P_{2} < P $.  Note that the bottom two plots correspond to the unstable cases where external servo is necessary to keep the system stable.}
\end{figure}

\begin{figure}[!h]
  \centering
\includegraphics[width=0.9\textwidth]{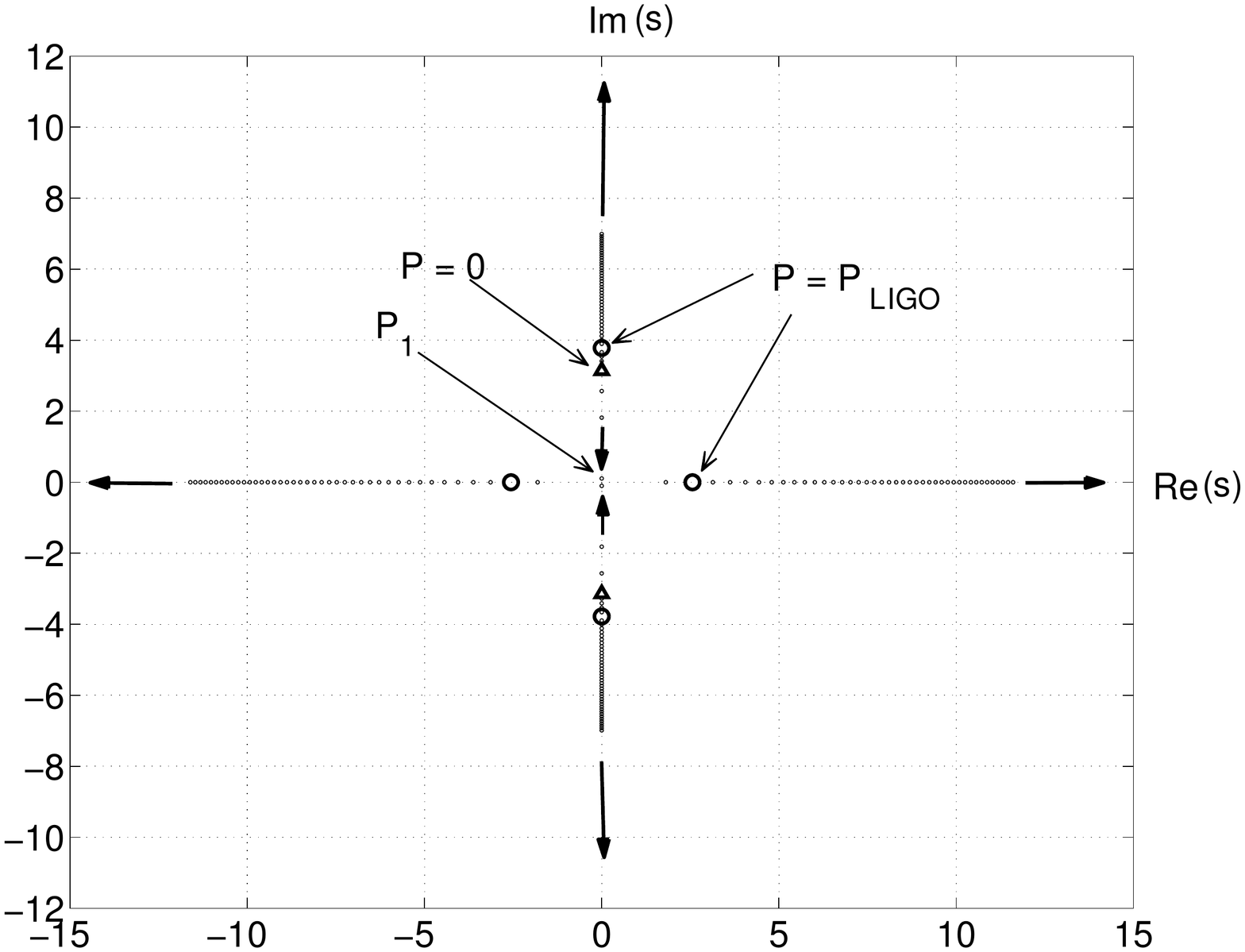} 
  \caption{Poles in the  s-plane at various powers, for the uncontrolled system. Triangles on the imaginary axis at $ P = 0 $ split into two directions: $ \omega_{+} $ moves away from the origin while $ \omega_{-} $ moves toward the origin.  After the pair for $ \omega_{-} $ hits the origin at $ P = P_{1} $, the poles move apart along the real axis.}
  \label{fig:poles_noctrl}
\end{figure}

\noindent When $ P = 0 $, the two mirrors are uncoupled.  The transfer function must be that for a simple torsion pendulum.  When $P$ is non-zero but sufficiently small, the two mirrors will be coupled and thus the Bode diagram will have two peaks associated with the two pairs of poles and one dip from the pair of zeros.  In the s-plane, the two pairs of poles at $ s = \pm i\omega_{0} $  denoted by triangles in Figure 6 split into non-degenerate pairs, one moving toward the origin and one moving away.  When $ P $ reaches $ P_{1} $, the pair associated with $ \omega_{-} $ reaches the origin, and for higher powers those poles move along the real axis.  When $ P $ reaches $ P_{2} $, above which a pair of zeros become imaginary, the dip corresponding to $ \omega_{z} $ will disappear from the Bode diagram. The circles in Figure 6 are the poles at the operating laser power in initial LIGO.  The existence of the pole in the right half plane indicates instability.  At $P>P_{1}$, initial LIGO would be unstable in the absence of an angular control system. 

\section{Measurement}

\begin{figure}[!h]
  \centering
\includegraphics[width=1.0\textwidth]{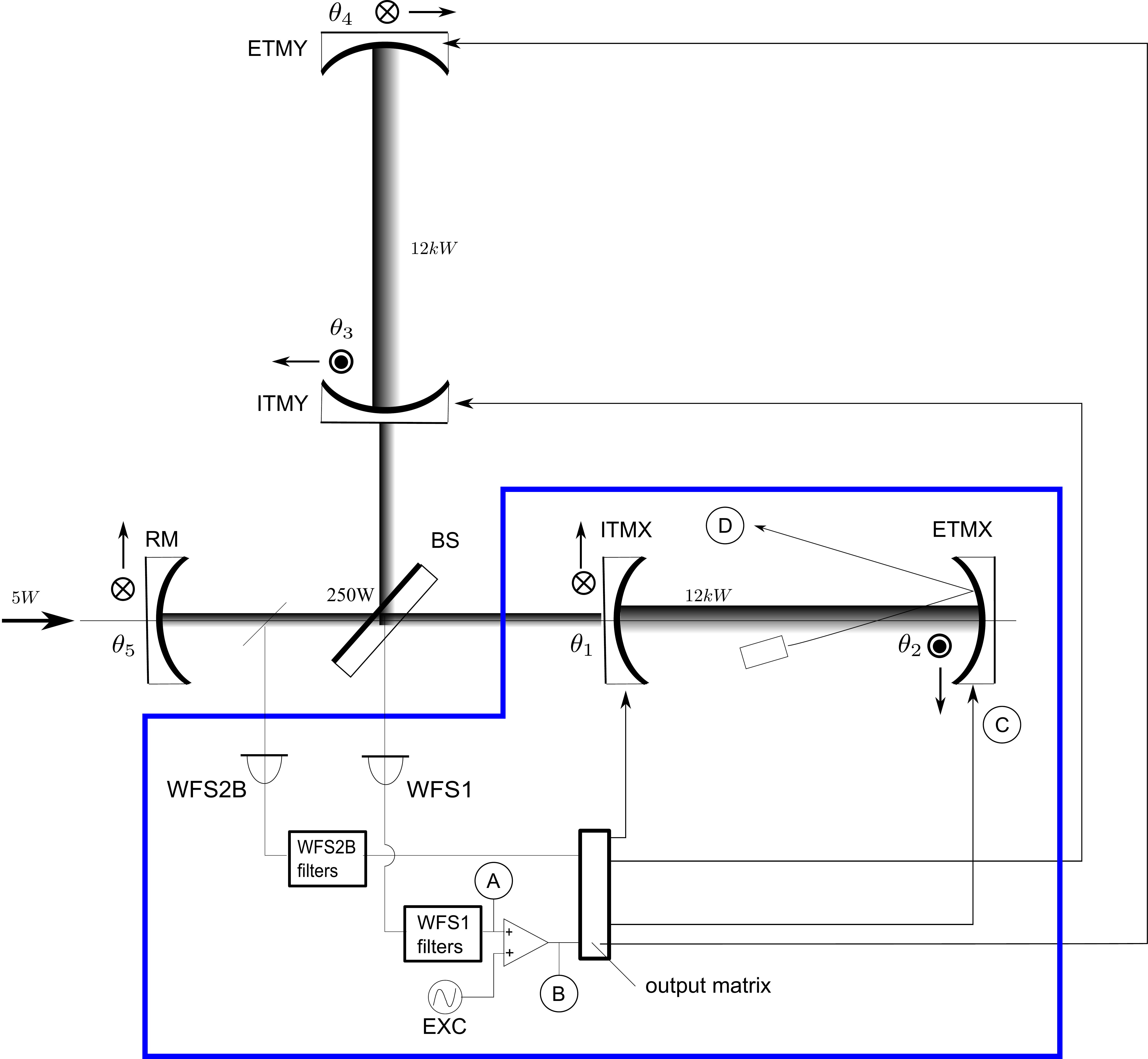} 
  \caption{Block diagram of our measurement scheme.  An excitation signal is injected into the WFS1 path that dithers ETMX and ETMY differentially, i.e., $ \Delta \theta_{ETM} = \theta_{2}-\theta_{4} $.  Both WFS1 and WFS2B are sensitive to the differential DOF ($ \Delta \theta_{ETM} $, $ \Delta \theta_{ITM} $), but we focus on the contribution to the x-arm.  The symbols next to each core optic show the rotation axes for pitch and yaw tilts.  A positive tilt angle follows a right-handed rotation about the axis.}
  \label{fig:extraction}
\end{figure}

To study the radiation pressure effect, we made a set of measurements on one of the arm cavities in the 4 km interferometer at  LIGO Hanford Observatory.  We wanted to study the regime of $P>P_{1}$, as described above, so all control loops (including angular controls) were engaged during the measurement.  The angular sensing and control system \cite{FGMShSiggZ, YaronNergisSigg, SiggNergis, FBGMORSiggZ} has ten degrees of freedom.  They are the pitch and yaw motion of the five core optics (the two pairs of mirrors: (ITMX, ETMX) and (ITMY, ETMY) in the long Fabry-Perot cavities, and the recycling mirror (RM) located upstream of the Michelson interferometer, see Figure \ref{fig:extraction}.)  These degrees of freedom are measured by the quadrant photodetectors called wavefront sensors, \cite{Anderson,MorrisonMRWard1} and are controlled by electromagnetic actuators attached to the backs of the mirrors.  We reduce the ten degees of freedom to two by looking at only two wavefront sensors which are sensitive to the differential degrees of freedom. (i.e., $ \Delta \theta_{ETM} = \theta_{2}-\theta_{4} $, and $ \Delta \theta_{ITM} = \theta_{1}-\theta_{3} $)\\

\noindent Besides these wavefront sensors for global angular sensing, there are sensors called optical levers \cite{KawamuraZucker}, which work only locally.  An optical lever consists of a diode laser and a quadrant photodiode which senses the position of the laser spot reflected by the mirror, and thus allows a measurement of the angle of the mirror.  We use the optical lever signal to monitor the response of the mirrors to angular excitation; this enables us to measure the transfer function $ H(s) $ that we investigated in the previous section.\\


\noindent In order to start the measurement, we first ensure that the interferometer is in its ordinary operating condition, where the two arm cavities and the recycling cavity are resonant, and the complete control system is engaged.  Then, we inject an excitation signal into one of wavefront sensors control path that results in dithering the end test masses differentially. During the excitation, we monitor the optical lever signal of the end mirror in the x-arm and a signal which goes into the driver of the actuator attached to the mirror.  Since the driver signal is proportional to the torque produced by the actuator, we are able to monitor the transfer function of the mirror angle $ D $ to torque $ C $ applied to the same mirror up to some overall gain.  The transfer function is the direct analog of the one we introduced in the previous section, i.e., the response of the ETMX mirror angle to the torque applied to the mirror.  In addition, we record signals from two points located just upstream $ A $ and downstream $ B $ of the excitation point in the wavefront sensors control loop.  This enables us to calculate the open loop transfer function of the control loop.  We here call the two transfer functions $ oplev(s) $ and $ olg(s) $ respectively.

\begin{eqnarray}
oplev(s) = \frac{D}{C}\\
olg(s) = \frac{A}{B}
\end{eqnarray}

\begin{figure}[!h]
  \centering
\includegraphics[width = 1\textwidth]{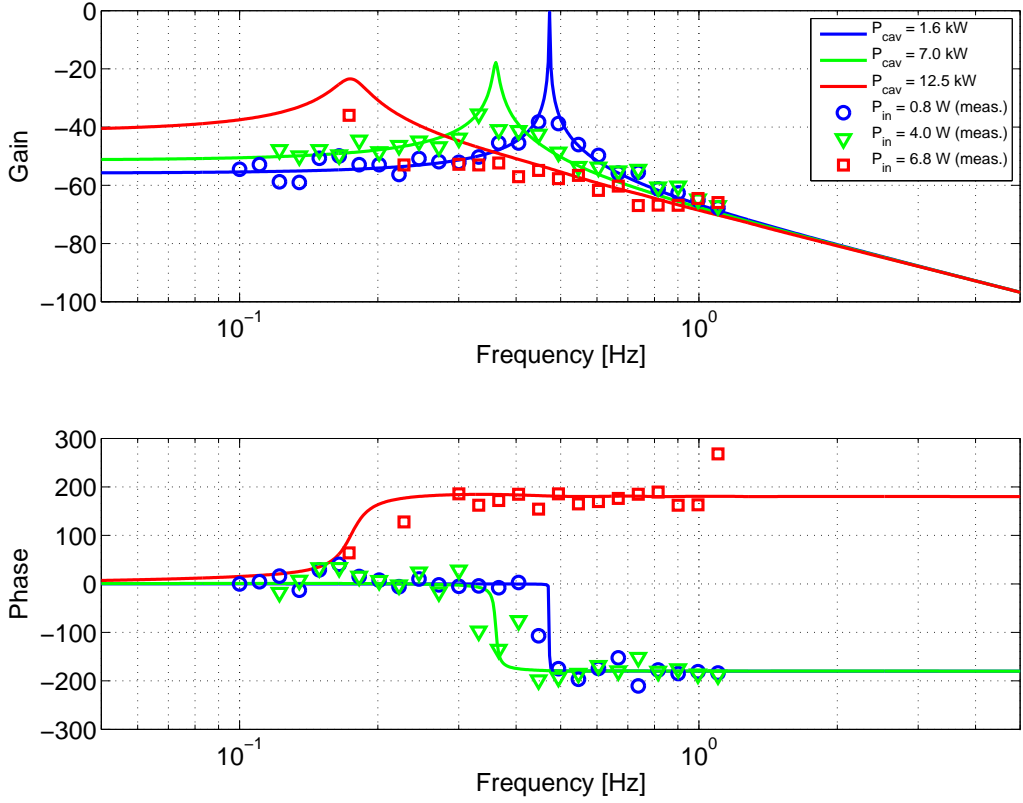}
  \caption{Bode plot of the optical lever transfer function, $ oplev(s) $. The input laser power levels are indicated as follows: 0.8 W :\textit{circles}, 4.0 W :\textit{triangles}, and 6.8 W :\textit{squares}.  The solid lines are the predictions of our mathematical model including the effect of control loops.}
  \label{fig:oplev}
\end{figure}

\begin{figure}[!h]
  \centering
\includegraphics[width = 1\textwidth]{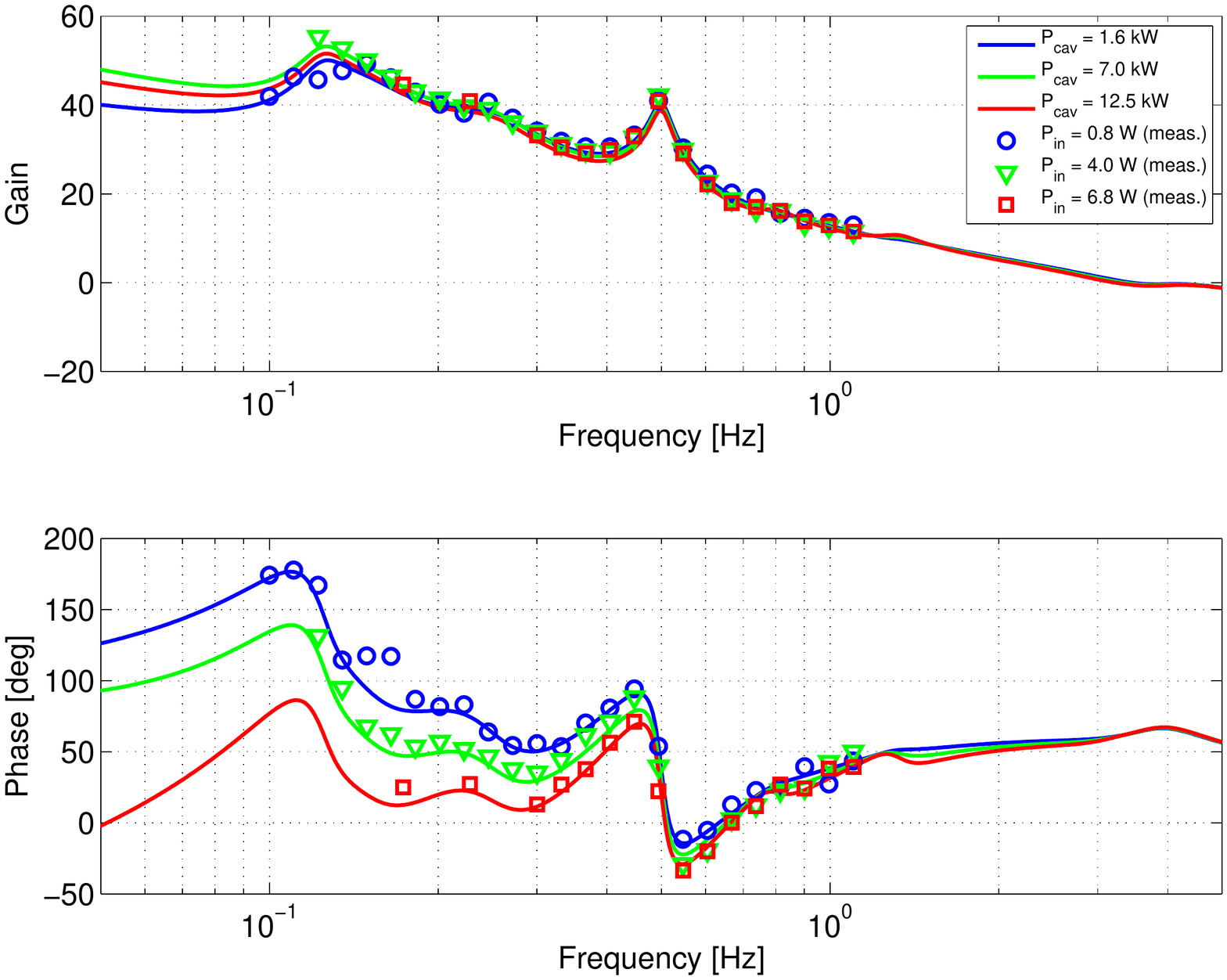}
  \caption{Bode plot of the open loop transfer function, $ olg(s) $.}
  \label{fig:olg}
\end{figure}

\noindent We made the measurement at three different input laser powers (0.8 W, 4.0 W, and 6.8 W.) \cite{HiroseKawabe}. In Figure \ref{fig:oplev}, the open symbols show the results of our measurements of the $ oplev(s) $ transfer function, while Figure \ref{fig:olg} shows our measurements of $ olg(s) $. (The solid lines are the predictions of our mathematical model including the effect of control loops, described in the next section.) \\

\noindent There are several surprising features in Figure \ref{fig:oplev}. Firstly, note that there is only one peak in the transfer function at each laser power, whose frequency decreases as the laser power goes up.  This is something that we did not expect based on the naive model (i.e., the model without any control loops).  Note also that, at the highest power level, a $ 180 $ degree phase lead was observed.  This was also surprising since it would usually indicate the system's instability.  \\

\section{Mathematical model}

These observations led us to construct a more complete model of the system that explicitly includes the dynamics of the angular control system. Figure 10 shows the block diagram.  The most important parts of the control loop are the sensing matrix (labeled as $ a,b,c,d $) and the control matrix ($ B,C,D $).  Each wavefront sensor measures a combination of the angle of the two mirrors (input and end mirrors); the signals are fed back to each mirror to minimize the deflection angle.  The sensing matrix is measured separately in advance by shaking each mirror; the control matrix is determined by inverting the sensing matrix.  Besides the wavefront sensors for global angular sensing, there is also a local loop called optical lever for each mirror, and a compensation loop to make the local loop invisible to the wavefront sensors.  All filter banks associated with the two wavefront sensors are built into the model.  The solid curves in Figures \ref{fig:oplev} and Figure \ref{fig:olg} show the transfer functions calculated by our models. The different colors in the plots denote different cavity laser power levels in the model.  They are blue for $ 1.6 $ kW, green for $ P = 7.0 $ kW, and red for $ P = 12.5 $ kW.  The agreement between the measurement and the model is very good.  The only adjustable parameter in the model is the cavity laser power. (related to the input power by an imperfectly determined factor.)  Knowing that the full interferometer has many more degrees of freedom than are modeled here, such agreement is especially gratifying. \\

\begin{figure}[!h]
  \centering
\includegraphics[width = 1\textwidth]{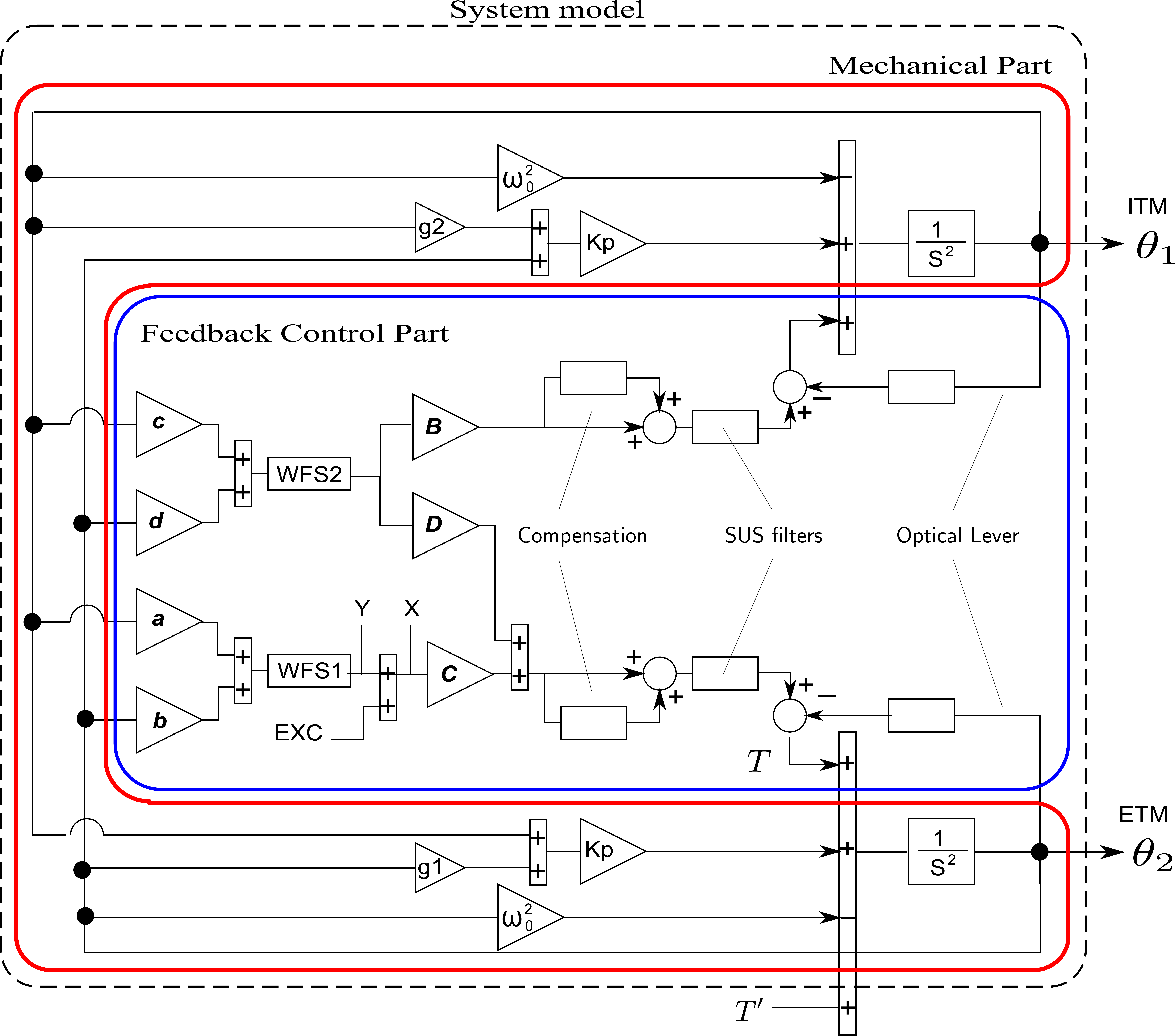} 
  \caption{Block diagram of interferometer arm with angular feedback control.  $ a,b,c,d $ and $ B,D,E $ are the sensing and control matrix, respectively.  Rectangular blocks stand for filter banks.  The SUS filters are ones related to the suspension system.  The factor $ K_{p} $ stands for the optical parameter  $ 2PL/\left[ cI(1-g_{1}g_{2})\right] $.}
\end{figure}

\noindent  There are several features worth noting in these transfer functions. Firstly, we can now understand why there is only one peak observed in the optical lever transfer function.  This resonance is associated with a pair of poles $ s = \pm i\omega_{-} $, but the control loops make the transfer function much more complicated than (4.1).  The reason why we only have one visible resonance is that with high loop gain as in initial LIGO, the pair of zeros $ s = \pm i \omega_{z} $ and the other pair of poles $ s = \pm i \omega_{+} $  asymptotically approach the same value and cancel each other. \\

\noindent Secondly, we can now understand why the phase of the optical lever transfer function at 6.8 W has a 180 degree phase lead.  As noted above, under the influence of a strong control loop, the pair of poles associated with the stable mode cancels the pair of zeros and only the pair of unstable poles survives.  In other words, when the control loop strongly suppresses the motion of the test masses, the in-loop transfer function $ oplev(s) $ looks as if it were a single mode unstable oscillator.  This is clearly observed in both our measurement and the model as the phase advancement in the plot.  This is evidence that the radiation pressure effect on optical resonators, predicted by Sigg and Sidles, really exists in the interferometer and that the system successfully operates in the regime where the instability would occur in the absence of control.  \\

\noindent Thirdly, the frequency associated with the peak shifts toward lower frequencies as laser power goes up.  This is understood as a radiation pressure effect, since the peak we observed corresponds to a pair of poles $ s = \pm i \omega_{-} $ and $ \omega_{-} $ decreases as shown in Figure 4. The frequency does not go down as rapidly with increasing power as in the naive model, since the control system enhances the effective restoring force on the ITM.\\  

\noindent Finally, we note that the radiation pressure effect is also observed in the phase of the open loop transfer function $ olg(s) $.

\section{Stability}

It is very important to know how much more power can be handled with the current control scheme for the upcoming higher power operation of the LIGO interferometers.  In particular, an upgrade to initial LIGO (called Enhanced LIGO) is now being developed.  It will have laser power roughly three times greater than initial LIGO, and will have an almost identical control system to that in initial LIGO.\\

\noindent When we discussed the stability of the system without control, a torque was applied to the system from 'outside'.  (We called it $ T $.)  However, with the control system in place, $T$ is no longer outside, but is included within the system.  Therefore, in order to judge the stability of the whole system, we need to consider another torque $ T' $, which is located truly outside of the whole system. (See Figure 10.) We will investigate the response of the system to such an outside torque

\begin{eqnarray}
H^{\star}(s) = \dfrac{\theta_{2}}{T'}.
\end{eqnarray}

\begin{figure}[!h]
  \centering
\includegraphics[width = 1\textwidth]{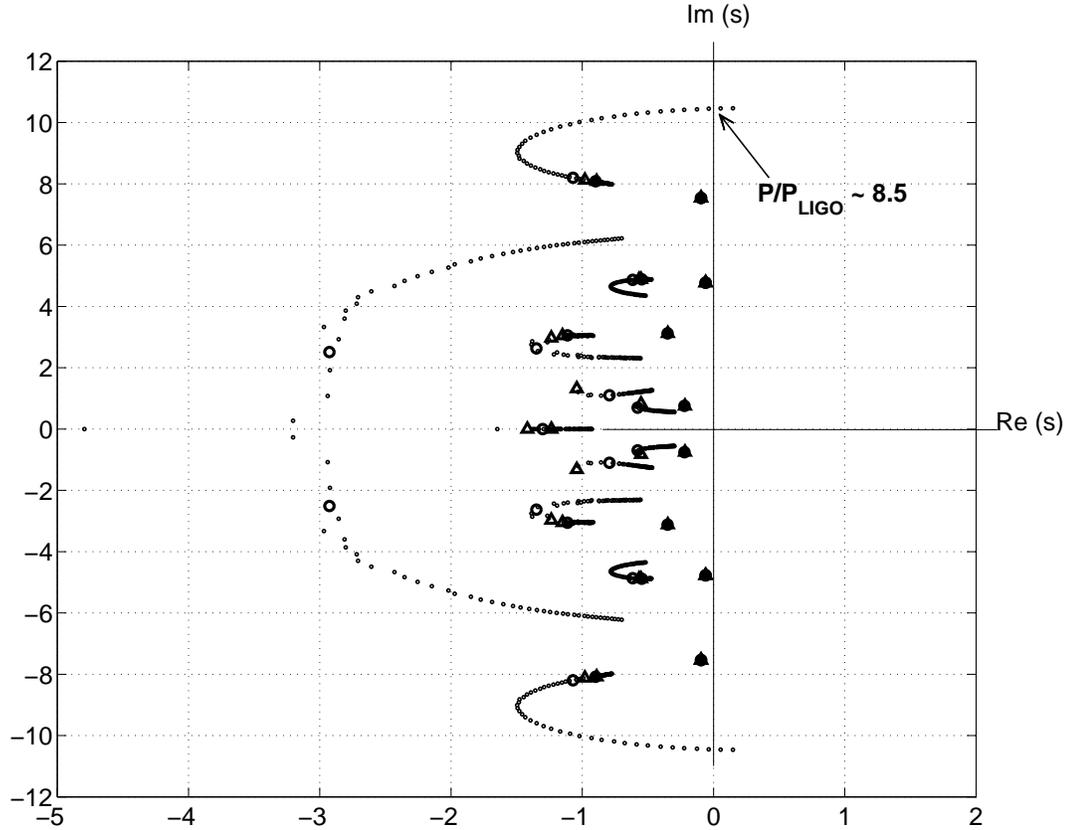} 
  \caption[Poles of the system with feedback control]%
  {Poles of the system with feedback control, with laser power from zero to $ 110 $ kW.  The power increment between points is $ 2.5 $ kW.  Triangles and circles are for $ P = 0 $, $ P = P_{LIGO} = 12.5 $ kW, respectively.}
  \label{fig:poles_ctrl}
\end{figure}

\noindent Figure \ref{fig:poles_ctrl} shows how the system poles move in the s-plane as a function of the cavity laser power.  We immediately notice the difference between Figure \ref{fig:poles_ctrl} and Figure \ref{fig:poles_noctrl}.  The circles, which indicate the systems poles at the full laser power level for initial LIGO, are now all in the left half plane, as expected.  According to our model, the system will remain stable until the cavity laser power reaches roughly $ P/P_{LIGO} = 8.5 $.  \\

\noindent This analysis was based on the degrees of freedom associated with differential motion between the mirrors in the interferometer's two arms.  The common mode degrees of freedom have lower unity gain frequencies in LIGO.  This means that the full initial LIGO system would probably not remain stable at powers as high as are predicted in our analysis.  A more complete analysis including the common mode degrees of freedom will be left for future work.  We also note that a new approach on this issue is being tested during the Enhanced LIGO commissioning process \cite{LisaMatt}.  It uses a control scheme built around the \textit{stiff} and \textit{soft} modal basis.\\

\noindent  Other workers in the field have also made studies of the radiation-pressure-induced angular instability in optical cavities. Sakata has demonstrated the effect in a specially-built apparatus at the National Astronomical Observatory of Japan \cite{Sakata}.


\section{Summary}

We observed the effect of radiation pressure on the angular control of the LIGO core optics at the 4 km interferometer at LIGO Hanford Observatory.  This is the first measurement of this effect performed on a full gravitational wave interferometer.  Only one of two angular modes survives with feedback control since the other mode is suppressed when the control gain is large enough.  A mathematical model was developed to understand the physics. It indicates that the system will remain stable at substantially higher power levels.

\section*{Acknowledgments}

We are grateful to Dr. Sam Waldman for useful discussions.  This research was supported in part by the National Science Foundation under grant PHY-0600259. LIGO was constructed by California Institute of Technology and Massachusetts Institute of Technology with funding from the National Science Foundation and operates under cooperative agreement PHY-0107417.  This article has LIGO document number LIGO-P0900086.

\end{document}